\documentstyle[11pt]{article}

\def\beq{\begin{equation}}
\def\eeq{\end{equation}}
\def\noi{\noindent}

\begin{document}
\hspace{10 cm}{\Large\bf LU TP 00-06}\\\vspace{1 cm}

\begin{center}
{\Large\bf Structure function of the nucleus in the perturbative QCD 
with $N_c\rightarrow\infty$\\
(BFKL pomeron fan diagrams)}\\\vspace{1 cm}

M.Braun \\ Department of High Energy physics,
 University of S.Petersburg,\\
198904 S.Petersburg, Russia
\end{center}
\vspace{1 cm}

\noi{\bf Abstract.}

Equation for the sum of BFKL pomeron fan diagrams 
is rederived by direct summation and  
solved 
numerically for rapidities $y\leq 50$. At high rapidities $y>20$
the resulting cross-sections for the scattering of a longitudinally
polarized $q\bar q$ pair on
the nucleus cease to depend on its transverse dimension and tend to a 
constant limit 0.1768 $R_A^2$, which corresponds to scattering of
a colour dipole on a black disk.
Thus the
unitarity is restored and the 
singularity in the $j$ plane is reduced to a simple pole at $j=1$. 
The nuclear
structure function at small $x$ behaves as $Q^2\ln(1/x)$.
The found gluon density has a
soliton-like form in the $\log k$ space: its form is close to Gaussian,
independent
of rapidity, the center
moving towards higher $\log k$ with a nearly constant velocity as
rapidity increases.

\section{Introduction}
In the framework of the colour dipole model of A.H.Mueller [1,2]
it follows that in the high-colour limit $N_c\rightarrow\infty$ the scattering 
on a heavy nucleus is exactly described by the sum of fan diagrams constructed 
of BFKL pomerons, each of them splitting into two [3]. This sum seems to be 
unitary by itself. It is
important 
that no splitting into three or more pomerons, as introduced in [4],
occurs,
although formally they give contributions of the same order.
Because of this fact, once the splitting vertex is known, construction of the 
amplitude for the interaction with the
nucleus becomes straightforward, reducing to summing BFKL pomeron fan diagrams.
This procedure has been well-known since the times of the old Regge-Gribov 
theory [5].

In the perurbative QCD a pioneering step was taken also many years ago 
 by L.Gribov, E.Levin and 
M.Ryskin, who summed fan diagrams in the double log approximation
and wrote their well-known non-linear GLR equation [6]. In the framework of
the BFKL dynamics the necessary tool for constructing fan diagrams is the
corresponding
triple pomeron vertex, which
was found in the colour diplole approach for $N_c\rightarrow\infty$
  by A.H.Mueller and 
B.Patel [2] and in the $s$-channel unitarity approach for any number 
of colours  by J.Bartels and M.Wuesthoff [7]. 
The equivalence of both results was shown in [8].

The equation for the sum of  BFKL fan
diagrams with this splitting vertex  was
written by I.Balitsky [9] in his original operator expansion formalism
and by Yu.Kovchegov [10] in the colour dipole framework (with a somewhat
unconventional form of the coupling to the target) .  
Its perturbative solution in the region of small non-linearity (outside 
the saturation region) was studied in [11]. Asymptotic estimates of the
solution were presented in [12].

In the present paper we first rederive the BFKL fan diagram equation by direct 
summation using the standard form of the pomeron-target coupling.  
It has a form of a simple (and elegant) evolution equation in 
rapidity $y$ for a wave function $\phi(y,q)$ in the momentum space:
\beq
\frac{\partial\phi(y,q)}{\partial y}=-H\phi(y,q)-\phi^2(y,q),
\eeq
where $H$ is the BFKL Hamiltonian for the so-called semi-amputated function
(a similar form was also obtained in [11]).

In spite of its tantalizing simplicity, Eq. (1) does not seem to allow for an 
analytical treatment except by perturbative methods, not valid
in the most interesting region of strong non-linear effects, or by
qualitative asymptotic estimates. The
bulk of this paper is correspondingly devoted to its 
numerical analysis. We numerically study evolution of the wave functuion in
 rapidity, starting from an appropriately chosen initial function. The results 
are then used to find the structure function of the nucleus at small $x=e^{-y}$
 and various virtualities $Q^2$. Our results show that for a heavy nucleus
the longitudinal part of
structure function saturates at $x\rightarrow 0$ to a universal function 
$F_{2L}^{(as)}(Q^2)$, independent of the nucleus atomic number but strongly 
dependent on $Q^2$ in the whole range of $Q^2\leq 10^5$ (GeV/c)$^2$ explored. 
In fact $F_{2L}^{(as)}(Q^2)$ is just proportional to $Q^2$. This reflects the
behaviour of the scattering cross
-section of a longitudinally polarized $q\bar q$ pair on the nucleus:
at very small $x$ it becomes
independent of both the pair size and $x$. The latter property indicates
that the unitarity is restored and the leading singularity in the complex
momentum $j$ is reduced to a simple pole at $j=1$. Note that the limiting
cross-section is in accordance with a picture in which the longitudinally
polarized photon with a certain probability splits into colour dipoles,
which then scatter on the nucleus as on a black disk.
For the leading transverse part of the structure function this probability
results infinite. Due to this fact both the cross-section and the structure
function continue to grow at $x\rightarrow 0$ approximately as $\ln (1/x)$.
These results are in agreement with predictions made in [11,12].

As to the gluon density,
we have found it to have a Gaussian shape as a function of $\xi=\ln k$
with a center at $\xi_0\propto y$. So with the growth of rapidity it propagates
towards higher momenta, practically preserving its form, very much like a
soliton
wave. Obvious limitations on computing time and memory
allow to follow its movement up to momenta not higher than of the order 
$10^{10}$ GeV/c. However we expect that this behaviour persists in the model
until arbitrary high values of momenta. As a result, at any fixed value of $k$
the density eventually goes to zero as $y\rightarrow\infty$. In this sense we
have the maximal saturation of the density possible.

The paper is organized as follows. In section 2 we present derivation of the 
BFKL fan diagram equation by direct summation.  
Section 3 is devoted to the numerical solution of Eq. (1). In Section 4 
the nuclear structure function and gluon density are calculated. Section 5 
contains our conclusions. In the Appendix
we compare our BFKL fan diagram equation with the Kovchegov's one [10]. 
\section{Fan diagram equation for the BFKL pomerons}

We start with a single scattering contribution to the forward ampitude
for the interaction of the projectile particle with a nucleus.
In the BFKL framework, at fixed impact parameter $b$, it has a well-known
form
\beq{\cal A}_1(y,b)=isg^4AT(b)\int d^2rd^2r'\rho(r)G(y,r,r')
\rho_N(r').\eeq
Here $\rho$ and $\rho_N$ are the colour densities of the projectile
and the target nucleon respectively. Function $G$ is the forwrad
BFKL Green function [13]:
\beq
G(y,r,r')=\frac{rr'}{32\pi^2}
\sum_{n=-\infty}^{+\infty}e^{in(\phi-\phi')}
\int_{-\infty}^{\infty}\frac{d\nu e^{y\omega(\nu)}}
{[\nu^{2}+(n-1)^{2}/4][\nu^2+(n+1)^2/4]}(r/r')^{-2i\nu},
\eeq
where $\phi$ and $\phi'$ are the azimuthal angles and
\beq
\omega(\nu)=2(\alpha_sN/2\pi)(\psi(1)-{\rm Re}\psi(1/2+i\nu))
\eeq
are the BFKL levels.
Due to  the azimuthal symmetry of the projectile colour
density one may retain only the term with zero orbital momenta
$n=0$ in (3). 
Separating the projectile part, the single scattering term may be
written in the form
\beq
{\cal A}_1(y,b)=2is\int d^2r\rho(r)\Phi_1(y,b,r),
\eeq
where
\beq
\Phi_1(y,b,r)=\frac{1}{2}g^4AT(b)\int d^2r'G(y,r,r')\rho_N(r').
\eeq

The double scattering contribution has been calculated in [14]. In the
limit $N_c>>1$ one finds
\[
{\cal A}_2(y,b)=-is\frac{g^2N_c}{4\pi^3}\int d^2r \rho(r)
\int_0^y dy_2 \int\prod_{i=1}^3 d^2r_i
\delta^2(r_1+r_2+r_3)\]\beq\frac{r_1^2\nabla_1^4}{r_2^2r_3^2}
G(y-y_2,r,r_1)\Phi_1(y_2,b,r_2)\Phi_1(y_2,b,-r_3).
\eeq
Presenting ${\cal A}_2$ in terms of $\Phi_2$ similarly to (5)
we find
\[
\Phi_2(y,b,r)=-\frac{g^2N_c}{8\pi^3}
\int_0^y dy_2 \int\prod_{i=1}^3 d^2r_i
\delta^2(r_1+r_2+r_3)\]\beq\frac{r_1^2\nabla_1^4}{r_2^2r_3^2}
G(y-y_2,r,r_1)\Phi_1(y_2,b,r_2)\Phi_1(y_2,b,-r_3).
\eeq

The whole set of fan
diagrams will be evidently summed by the equation
which is graphically illustrated in Fig. 1:
\[
\Phi(y,b,r)=\Phi_1(y,b,r)
-\frac{g^2N_c}{8\pi^3}
\int_0^y dy_2 \int\prod_{i=1}^3 d^2r_i
\delta^2(r_1+r_2+r_3)\]\beq\frac{r_1^2\nabla_1^4}{r_2^2r_3^2}
G(y-y_2,r,r_1)\Phi(y_2,b,r_2)\Phi(y_2,b,-r_3).
\eeq
The impact parameter $b$ appears here only as a parameter and the
dependence on it will be 
implicit in the following. In terms of $\Phi$ the total forward scattering
amplitude on the
nucleus  will be given at fixed $b$ by
\beq
{\cal A}(y,b)=2is\int d^2r\rho(r)\Phi(y,b,r).
\eeq

One can rewrite Eq. (9) as an evolution equation in $y$.
To this end we  present $\Phi$
as an integral over $\nu$'s similar to (3):
\[
\Phi(y,r)=\int d\nu r^{1-2i\nu}\Phi(y,\nu).
\]
Then from (9) we find 
\[
\Phi(y,\nu)=\frac{g^4}{64\pi^2(\nu^2+1/4)^2}e^{\omega(\nu)y}
\int d^2 r'r'^{1+2i\nu}\rho(r')AT(b)-\]\[
\frac{g^2N_c}{32\pi^5(\nu^2+1/4)^2}e^{\omega(\nu)y}
\int_0^y dy_2 e^{-\omega(\nu)y_2}\int\prod_{i=1}^3 d^2r_i
\delta^2(r_1+r_2+r_3)\]\beq\frac{r_1^2\nabla_1^4}{r_2^2r_3^2}
r_1^{1-2i\nu}\Phi(y_2,r_2)\Phi(y_2,-r_3).
\eeq
Multiplying both parts by $e^{-\omega(\nu)y}$ and taking a derivative in
$y$
we obtain
\[
\left(\frac{\partial}{\partial y}-\omega(\nu)\right)\Phi(y,\nu)=
-\frac{g^2N_c}{32\pi^5(\nu^2+1/4)^2}
\int\prod_{i=1}^3 d^2r_i
\delta^2(r_1+r_2+r_3)\]\beq\frac{r_1^2\nabla_1^4}{r_2^2r_3^2}
r_1^{1-2i\nu}\Phi(y,r_2)\Phi(y,-r_3).
\eeq
Returning to the $r$ space we take into account that
\beq \omega(\nu)r^{1-2i\nu}=-\hat{H}r^{1-2i\nu},\eeq
where $\hat{H}$ is the BFKL Hamiltonian [13]. Then  (12) transforms into
\beq
\left(\frac{\partial}{\partial y}+\hat{H}\right)\Phi(y,r)=
-\frac{g^2N_c}{8\pi^3}
\int\prod_{i=1}^3d^2r_i
\delta^2(r_1+r_2+r_3)\frac{r_1^2\nabla_1^4}{r_2^2r_3^2}
G(0,r,r_1)\Phi(y,r_2)\Phi(y,-r_3).
\eeq
Using
\beq
\frac{r_1^2\nabla_1^4}{r^2}G(0,r,r_1)=
\frac{1}{2\pi^2rr_1}
\sum_{n=-\infty}^{+\infty}e^{in(\phi-\phi_1)}
\int_{-\infty}^{\infty}d\nu 
(r/r_1)^{-2i\nu}=\delta^2(r-r_1),
\eeq
we simplify (14) to a form
\beq
\left(\frac{\partial}{\partial y}+\hat{H}\right)\Phi(y,r)=
-\frac{g^2N_c}{8\pi^3}
\int\prod_{i=2}^3d^2r_i
\delta^2(r+r_2+r_3)\frac{r^2}{r_2^2r_3^2}
\Phi(y,r_2)\Phi(y,-r_3).
\eeq

The initial condition is determined from (9) to be
\beq
\Phi(y,r)_{y=0}=\Phi_0(r)
\eeq
with
\beq
\Phi_0(r)=\frac{1}{2}g^4AT(b)\int d^2r'G(0,r,r')\rho(r').
\eeq
Except for the initial condition and different variables, Eq. (16)
coincides with the one
constructed by Kovchegov in the colour dipole approach in [10] (see
Appendix).

Now we go to function
\beq
\phi(y,r)=\frac{1}{r^2}\Phi(y,r)
\eeq
and pass to the momentum space. For $\phi(y,q)$ 
Eq. (16) reads
\beq
\left(\frac{\partial}{\partial y}+\frac{1}{r^2}\hat{H}r^2\right)\phi(y,q)
=-\frac{g^2N_c}{8\pi^3}
\phi(y,q)\phi(y,q),
\eeq
where we have used that $\phi(y,q)$ in fact depends only on $|q|$,
which follows from the initial conditon
\beq
\phi_0(y,q)_{y=0}=\int\frac{d^2r}{r^2}e^{-iqr}\Phi_0(r)
\eeq
with $\Phi_0$  given by (18) and depending only on $|r|$ due
to the azimuthal summetry. 

The Hamiltonial $r^{-2}\hat{H}r^2$ which
appears on 
the left-hand side is  the standard forward Hamiltonian for 
the semi-amputated functions 
\beq
\frac{1}{r^2}\hat{H} r^2=H=\frac{g^2N_c}{4\pi^2}
[\ln q^2+\ln r^2 -2(\ln 2 +\psi(1))].
\eeq
Indeed the eigenfunctions of $\hat{H}$
$\Phi_{n,\nu}(r)=r^{1-2i\nu}e^{in\phi}$ go over into the
eigenfunctions of $H$ after division by $r^2$. 

This brings us to nearly the final form of our equation
\beq
\left(\frac{\partial}{\partial y}+H\right)\phi(y,q)
=-\frac{g^2N_c}{8\pi^3}\phi^2(y,q).
\eeq
It remains only to appropriately rescale $H$, $\phi$ and $y$ to obtain
the BFKL fan diagram equation in the final form (1). We introduce
\beq
H=\frac{g^2N_c}{4\pi^2}\tilde{H},\ \
\tilde{y}=\frac{g^2N_c}{4\pi^2}y,\ \ {\rm and}\ \ 
\phi(\tilde{y},q)=2\pi\tilde{\phi}(\tilde{y},q),  
\eeq
where
\beq
\tilde{H}=\ln r^2+\ln q^2-2(\ln2+\psi(1)).
\eeq
The equation for $\tilde{\phi}$ takes the form (1)
\beq
\left(\frac{\partial}{d\tilde{y}}+\tilde{H}_0\right)\tilde{\phi}
(\tilde{y},q)=-\tilde{\phi}^2(\tilde{y},q)
\eeq
with the initial condition
\beq
\tilde{\phi}(\tilde{y},q)_{\tilde{y}=0}=
\frac{g^4AT(b)}{4\pi}\int d^2rd^2r_1\frac{1}{r^2}e^{iqr}
G(0,r,r_1)\rho_N(r_1).
\eeq

\section {Numerical solution of the BFKL fan diagram equation}
To solve numerically Eq. (26) we first transform the BFKL Hamiltonian
$H$ to more convenient variables. In the momentum space
one can write action of $H$ on a function $\phi(q)$ as
\beq
H\phi(q)=-2\int_0^{\infty}kdk\Big[\frac{\phi(k)-(q^2/k^2)\phi(q)}
{|q^2-k^2|}+\frac{q^2}{k^2}\frac{\phi(q)}{\sqrt{4k^4+q^4}}\Big].
\eeq
We transform it to variables $u,v$ which take values in [0,1]:
\beq
 q=\exp[(M_1+M_2)u-M_1],\ \  k=\exp [(M_1+M_2)v-M_1],
\eeq
where $q,k$ are in GeV/c and $M_{1(2)}$ is a lower(upper)
integration limit in $\ln k$.
Then (28) goes into
\beq
H\phi(u)=-2(M_1+M_2)\int_0^1dv\Big[\frac{\phi(v)-f(u-v)\phi(u)}
{|f(u-v)-1|}+\frac{\phi(u)}{\sqrt{1+4f^{-2}(u-v)}}\Big],
\eeq
where
\beq 
f(u)=\exp[(M_1+M_2)u].
\eeq

Now we discretize the interval [0,1] in $u$ and $v$ into $n$ equidistant
points $u_0,u_1,...u_n$ and $v_0,v_1,...v_n$ to convert the evolution 
equation (26) into a set of $n$
ordinary 1st order non-linear differential equations  in $y$.
This set of equations can be solved by standard methods. We used
the simplest 2nd order Runge-Kutta algorithm. The standard values of 
$n$ and of the number of iterations to evolve in two units 
of $\tilde{y}$ were 800. We have studied evolution of $\phi(y,q)$
from the initial value at $y=0$ up to $\tilde{y}=10$. The (fixed) 
value of the strong coupling $\alpha_s=g^2/4\pi$ has been chosen
to be 0.2. With this choice our maximal rapidity is around 50. 

An evident difficulty which one meets are the values of $H\phi$ at
endpoints $u_0$ and $u_n$, at which the introduced cutoffs make the
results not reliable. To overcome this difficulty we calculated
$H\phi$ at these points by extrapolation from the neighbouring
points. The stability of this procedure was checked by comparing
the results for double and quadruple values of $n$.

The initial function is determined by the 
colour density of the nucleon $\rho_N(r)$ according to (27).
To simplify our calculations we have taken the Yukawa form for
$\rho_N(r)$:
\beq
\rho_N(r)=\frac{\mu}{2\pi}\frac{e^{-\mu r}}{r},
\eeq
where $\mu$=1/0.7 fm has a meaning of the inverse nucleon radius.
This choice may look a bit arbitrary, but our results show that with the
growing rapidity the system quickly forgets not only the form but even
the absolute magnitude of $\rho_N$, so that the solution becomes
independent of the initial function and 
governed only by the internal dynamics of Eq. (26) itself. 

Doing the integrations over $r$ and $r_1$ in (27) and over $\nu$ inside 
the BFKL Green function (3) we find for $q>\mu$
\beq
\tilde{\phi}_0(q)_{q>\mu}=B
\frac{\mu^2}{q^2}
\Big\{1-\frac{\mu}{q\sqrt{\pi}}
\sum_{n=0}(-1)^n\left(\frac{\mu^2}{q^2}\right)^n
\frac{\Gamma(1/2+n)}{(n+3/2)^2}\Big\}
\eeq
and for $q<\mu$
\[
\tilde{\phi}_0(q)_{q<\mu}=B
\Big\{\frac{1}{4}[(2\ln\frac{q}{\mu}+\psi(3/2)-\psi(1)-1)^2+
\psi'(3/2)+\psi'(1)+1]-\]\beq \frac{q^2}{\sqrt{\pi}\mu^2}
\sum_{n=0}(-1)^n\left(\frac{q^2}{\mu^2}\right)^n
\frac{\Gamma(5/2+n)}{(n+2)(n+1)^3}\Big\},
\eeq
where the dimensionless coefficient
\beq
B=\frac{g^4}{16\pi}\frac{AT(b)}{\mu^2}
\eeq
carries all the information about the nucleus. Its maximal value 
with $\alpha_s=0.2$ is about 0.12 for the central scattering on lead
($b=0$).
The asymptotic behaviour of
$\phi_0(q)$ at $q\rightarrow\infty$ and $q\rightarrow 0$ is governed by 
the first terms in (33) and (34), which come from the poles of the BFKL
Green 
functions at $\nu=\pm i/2$:
\[
\phi_0(q)\sim 1/q^2,\ \ q\rightarrow\infty,
\]\beq
\phi_0(q)\sim\log^2q,\ \ q\rightarrow 0.
\eeq
We have checked that our results parctically do not change if the
complicated function which multiplies $B$ in (33) and (38) is
substituted just by $\ln^2(q/(\mu+q)$ with the same asymptotic behaviour
(36).

The initial function 
depends on
the form of the nucleus profile function $T(b)$. For our numerical 
calculation we have chosen  
\beq
T(b)=2\sqrt{R_A^2-b^2}/V_A
\eeq
corresponding to a finite nucleus of radius $R_A$ 
and volume $V_A$ with a constant density. We have taken $R_A=A^{1/3}R_0$ with
$R_0=1.2$ fm.

Our results for $\phi(q)$ for two values of $B=0.12$ and 0.02,
corresponding to a central and very peripheral  collisions off lead,
respectively, are presented in Figs 2,3. One observes that 
 the solution $\phi$ at rapidities $\tilde{y}>2$ 
evolves to
a very simple and universal form, practically independent of
the initial function. Crudely speaking it linearly falls with $\xi=\ln k$
until it meets the $x$-axis wherefrom it stays equal to zero. The slope of the
falling part is exactly equal to 1, so that very crudely
\beq
\phi(\xi)=\xi_1(y,b)-\xi,\ \ {\rm for}\ \ \xi<\xi_1(y,b),\ \ \ 
\phi(\xi)=0,\ \ {\rm for}\ \ \xi>\xi_1(y,b).
\eeq
The value of $\xi_1(y,b)$ and hence the interception point with the $x$ axis
grow  linearly with $y$, so that with the growth of $y$ the
picture simply shifts to the right.
In reality  the curve for $\phi$  of course has no break:
the two straight lines of which it is formed join
smoothly in the vicinity of $\xi_1$. As will be clear later, the
physically important region is precisely this vicinity, where $\phi$ is
not trivial.

The results for $\phi(k)$ do not depend on the chosen cutoffs, provided
they are taken to cover the region around the interception point $\xi_1$.
Otherwise the evolution stops as soon as $\xi_1$ touches the upper cutoff
$M_2$. So if one wants to study evolution up to high values of $y$
the upper cutoff should be taken correspondingly high. In our calculations
we chose $M_1=10$ and $M_2=20$, having verified that further
raising of either $M_1$ or $M_2$ does not change the results.
 
In conclusion we find that $\phi$ does not
 possess
any finite limit as $y\rightarrow\infty$
so that Eq. (26) does not lead to any saturation of the wave function
$\phi$ at high rapidities, contrary to naive expectations.
However in the next chapter we shall see, that such saturation indeed occurs 
for
physical quantities. The point is that function $\phi$ by itself has no
physical meaning. It is its derivatives which matter. 

\section{Nuclear structure function and the gluon density}

The nuclear structure function is obtained in the standard manner as
\beq
F_2(x,Q^2)=\frac{Q^2}{\pi e^2}(\sigma_T+\sigma_L),
\eeq
where $\sigma_{T,L}$ are the total cross-section for the scattering 
on the nucleus of a virtual photon with transversal (T) or longidunal (L) 
polarization. Both cross-sections can be found from the imaginary part of
the forward scattering amplitude (10). In  terms
of $\tilde{\phi}$ we find
\beq
\sigma_{T,L}=4\pi\int d^2b d^2r\rho_{T,L}(r)r^2 \tilde{\phi}(y,r,b).
\eeq
Here we explicitly indicated the dependence of $\phi$ on the impact
parameter;
$\rho_{T,L}(r)$ are the well-known colour densities of the
virtual photon split into a $q\bar q$ pair (see e.g. [15]). With
massless quarks
\beq
\rho_T(r)= \frac{e^{2}N_cZ^2}{8\pi^3}
\int_{0}^{1}d\alpha
(\alpha^{2}+(1-\alpha)^{2})\epsilon^{2}{\rm K}_{1}^{2}
(\epsilon r))
\eeq
and 
\beq
\rho_{L}(r)= \frac{e^{2}N_cZ^2}{2\pi^3}Q^2
\int_{0}^{1}d\alpha
\alpha^2(1-\alpha)^2{\rm K}_{0}^{2}(\epsilon r),
\eeq
where
$\epsilon^2=Q^{2}\alpha (1-\alpha)$ and $Z^2$ is a sum of squares of
quark electric charges in units $e$.

Passing to momentum space we find
\beq
\sigma_{T,L}=4\pi \int d^2b
\frac{d^2q}{(2\pi)^2}\tilde{\phi}(q,y,b)w_{T,L}(q),
\eeq
where 
\beq
w_{T,L}(q)=\int d^2r r^2\rho_{T,L}(r)e^{iqr}.
\eeq
Straightforward calculation leads to the following expressions for
$w_{T,L}(q)$. For the transverse density one finds\footnote
{An error in this formula  lead to some erroneous
conclusuions about the $y$-behaviour of the structure function
in the original version} 
\beq
w_T(q)= \frac{e^{2}N_cZ^2}{8\pi^2}
\int_{0}^{1}d\alpha
(\alpha^{2}+(1-\alpha)^{2})\nabla_q^2[(q^2/2
+\epsilon^2)J(q,\epsilon)],
\eeq
where
\beq
J(q,\epsilon)=
\frac{2}{q\sqrt{q^2+4\epsilon^2}}\ln\frac{\sqrt{q^2+4\epsilon^2}+q}
{\sqrt{q^2+4\epsilon^2}-q}.
\eeq
The longitudinal density $w_{L}$ is given by the same expression
with substitutions
\[\alpha^{2}+(1-\alpha)^{2}\rightarrow \alpha(1-\alpha),\ \ 
q^2/2+\epsilon^2\rightarrow -4\epsilon^2\]

With the found numerical values for the function $\phi$ in the range
$0\leq\tilde{y}\leq10$ we evaluated the nuclear structure function of lead 
($A=207$) for various $Q^2$ between 3  and 10$^5$ (GeV/c)$^2$.
The results are most instructive  for the the cross-section 
$\sigma_{L}$ for the scattering of a longitudinally polarized
virtual photon on the nucleus.
In Figs. 4,5 we present it (with $e^2\rightarrow 1$) as a 
function of $y$ at fixed $Q^2$ and as a function
of $Q^2$ at fixed $y$ respectively (in GeV$^{-2}$).
From Fig. 4 one clearly sees saturation in rapidity: for any value of $Q^2$
the cross-section tends to the same limit of $228.8$ (GeV/c)$^{-2}=
0.1768 R_A^2$ as $\tilde{y}$ goes beyond 5.
The resulting constant cross-section is evidently 
consistent with the unitarity restrictions. In terms of the complex
angular momentum $j$ our results indicate that the original cut at
$j>1$ is reduced to a simple pole at $j=1$.
 Fig. 5 illustrates an unusual behaviour in
$Q^2$
which sets in at high $y$: instead of going down as $1/Q$ in the standard
BFKL approach, the cross-section becomes independent of $Q^2$ to a very
high precision.

These results for the longitudinal cross-section can be conveniently
interpreted in terms of scattering of colour dipoles off the nucleus.
The density $\rho_L(r)$, appropriately normalized, can be interpreted
as a probability distribution for the longitudinal photon to split into
colour dipoles of transverse dimension $r$. The normalization factor
\[ D=\int d^2r \rho_L(r)=\frac{e^2N_cZ^2}{12\pi^2}=0.028145\,e^2\]
can be considered as a total probability for the 
photon to split into dipoles. Then the dipole-nucleus cross-section is
found by dividing $\sigma_L$ by factor $D$, which gives
8131.1 (GeV/c)$^{-2}$
independent of $Q^2$, that is, of the dipole dimension. This value
exactly equals $2\pi R_A^2$, corresponding to scattering off
a black disk.

The transverse part of the structure function does not admit this
interpretation, due to the fact that $\rho_T(r)$ is not
normalizable. As a result both the cross-section and the  structure
function do not saturate at large $y$ but continue to grow nearly linearly
in $y$. This is illustrated
in Figs. 6,7 where the total structure function (including the
much smaller longitudinal part) is shown as a function of
$x=e^{-y}$ and $Q^2$ respectively.

We finally come to the gluon density. Although, strictly speaking it is not a
physical quantity, its properies have been much discussed recently
in connection with its saturation for the large nucleus [16].
It can be related to our function $\phi$ via the standard expression
for the structure function in its terms (see e.g [14])
\beq
F_2(x,Q^2)=
\frac{g^2Q^2}{\pi^3Ne^2}\int d^2b d^2r[\rho_T(r)+\rho_L(r)]
F(x,r,b),
\eeq
where
\beq
F(x,r,b)=\int\frac{d^2k}{(2\pi)^2k^4}k^2
\frac{\partial xG(x,k^2,b)}{\partial k^2}
\left(1-e^{-ikr}\right)\left(1-e^{ikr}\right)
\eeq
and $\int d^2b (\partial xG(x,k^2,b)/\partial k^2)$ is up to a
factor 
the gluon density in the momentum space:
\beq
\frac{\partial N(l)}{\partial^2 l}=\frac{1}{\pi}
\frac{\partial N(l)}{\partial l^2}=\frac{1}{\pi}
\int d^2b\frac{\partial xG(x,l^2,b)}{\partial l^2}.
\eeq
In the following we shall study the double density in momentum and
impact parameter, which is just $\partial xG(x,k^2,b)/\partial k^2$.
Comparing (47), (48) with the corresponding expression in terms of $\phi$,
which follows from (39) and (40) we find a relation
\beq
F(x,r,b)=\frac{4 N}{\pi g^2}r^2\tilde{\phi}(\ln\frac{1}{x},r,b).
\eeq
Taking a Fourier transform of (48) and neglecting the term proportional to
$\delta^2(k)$ we  obtain
\beq
\frac{\partial xG(x,k^2,b)}{\partial k^2}=
\frac{2 N}{\pi g^2}k^2\nabla_k^2\tilde{\phi}(\ln\frac{1}{x},k,b).
\eeq
This is the desired relation between our function $\phi$ and the
gluon density in the combined momentum and impact parameter space. 

Applying $k^2\nabla_k^2$ to the found function $\phi$  we thus find the gluon 
density up to a trivial numerical factor evident from (51)( $\sim 0.76$
with $\alpha_s=0.2$). Function $h(k)=k^2\nabla_k^2\tilde{\phi}(y,k,b)$ 
for
different values of $y$ and $B=0.12$ and 0.02 is shown in Figs. 8,9.  
Its form at different $y$ results quite remarkable. As one observes, 
at any given rapidity the  found density has the same, roughly Gaussian
shape in variable $\xi=\ln
k$,
centered at the point $\xi=\xi_0(y)$ very near to  $\xi_1$ at 
which the straight line (38) crosses the $x$-axis (see Section 2). With
the growth of $y$  the distribution
 moves to the right with a nearly constant velocity practically preserving
its form. Approximately the distribution can be described by 
\beq
h(k)=h_0 e^{-a(\xi-\xi_0(y))^2},
\eeq
where $h_0$ and $a$ are practically independent of $y$ and
$\xi_0(y)$ linearly grows with it:
\beq
h_0\simeq 0.3,\ \ a\simeq 0.3\ \ \xi_0(y)=\xi_{00}+2.23 \tilde{y}.
\eeq
The only quantity which clearly depends on the initial distribution is
the starting position $\xi_{00}$, so that for different initial functions
the picture in Figs. 8 and 9  shifts along the $\xi$ axis as a whole.

Evidently at a given value of $k$ the density stays always limited, 
irrespective of the form of the initial distribution (and on the atomic 
number $A$, in particular). In this sense we have saturation as discussed 
in [16]. However with the growth of $y$ the
strongly peaked density moves away toward higher values of $k$ so that the 
density at a fixed point tends to zero at  high values of rapidity.
We thus have  ``supersaturation'': with $y\rightarrow\infty$ the gluon
density
at an arbitrary finite momentum tends to zero.

Comparing Figs. 8 and 9 one can see how the memory about the initial
distribution (except for $\xi_{00}$) is 
gradually erased
in the course of the evolution in $y$.  For a very peripheric collison off
lead ($B=0.02$) at the initial stages of the evolution the density is
correspondingly much smaller than for a central collision ($B=0.12$).
However already at $\tilde{y}=2$ the form of the distribution
is practically indistinguishable from the central collision, 
the only remaining difference being the shift along the $\xi$ axis.

\section{Conclusions}
The BFKL fan diagram equation has been solved numerically in the
large range of rapidities up to $y=50$.
 The main results are the following.

The idea that the fan diagrams themselves satisfy the unitarity condition
has been supported by the fact that the found cross-sections for the
scattering of a $q\bar q$ pair off the nucleus tend to a constant value
at high rapidities.
Since the found cross-sections
do not rise with energy, the
leading $j$-plane singularity turns out to be a simple pole
at $j=1$.
The limiting cross-sections prove to be universal: they do not depend
on $Q^2$, that is, on the transverse dimension of the $q\bar q$ pair,
nor on the initial colour distribution inside the nucleon.
Their dependence on the target nucleus thus reduces to a scale
factor $R_A^2$. Physically they correspond to scattering of
a colour dipole off a black disk.

The nuclear structure function does not saturate at high rapidities,
due to the singularity of the transverse distribution
$\rho_T(r)$ at $r=0$, which
makes it non-normalizable.  At large $y$ it continues to grow nearly linearly
in $y$.

These results fully agree with predictions made in [11] on the basis of
the found perturbative solution of the BFKL fan diagram equation
and asymptotic estimates made in [12].

A completely novel result concerns the gluon density of the nucleus.
At sufficiently  high rapidities, greater than 10, the gluon density
aquires a form of the soliton wave in $y-\ln k$ space, which, with the
growth of $y$, moves
along the $\ln k$ towards greater $k$ preserving its nearly Gaussian shape.
Thus at any finite $k$ the  gluon density eventually goes to zero at
high enough $k$

Finally we have to stress that all these properties begin to be clearly 
visible only at very high rapidities and momenta: $y>10$ and
$k> 100$ GeV/c with  $\alpha_s=0.2$. With smaller $\alpha$'s
these values grow correspondingly.

\section{Acknowledgements}
The author is grateful to Dr.G.P.Vacca who draw his attention to a
possibility to simplify the BFKL fan diagram equation.
He is also grateful to Prof. Bo Andersson for his interest in this
work and stimulating discussions. He thanks the Department for Theoretical
Physics of the Lund University for hospitality during his stay in Lund,
where this work was completed.
The author is finally most thankful to Dr. Yu.Kovchegov, fruitful
discussions with whom helped to find and correct an error in the expression
for $w_T$ (Eq. (45)) in the first version of this paper.

\section{Appendix. Colour dipole approach}
To compare our BFKL fan diagram equation to that of Kovchegov [10] we 
present here a short derivation of Eq. (1) from the colour dipole
approach, which will make clear the difference between the two
equations.

In the colour dipole approach the single scattering term (2) is presented
in the form
\beq
{\cal A}(y,b)=-2isAT(b)\int d^2r\rho(r)\int d^2r_1n_1(r,r_1,y)\tau(r_1),
\eeq
where
\beq
\tau(r_1)= -\frac{1}{2}g^4\int d^2r'G(0,r_1,r')\rho_N(r')
\eeq
and $n_1(r,r_1,y)$ is a single dipole density at rapidity $y$
introduced by A.H.Mueller; $r_1$ and $r$ are the  dipole
lengths at rapidity $y$ and at $y=0$ respectively, so that
\beq
n_1(r,r_1,y)_{y=0}=\delta^2(r-r_1).
\eeq
(Note that as in [10] our $n_1$ is A.Mueller's one divided by $2\pi
r_1^2$.)

 To introduce the multidipole densities
A.H.Muller constructed a generating functional $Z(r_1,r_0,y|u)$,
where $u=u(\rho_i,\rho_f)$ is a function of two dipole endpoints
in the transverse space (which, for brevity, Iwe denote
 by a single symbol $\rho$ for).
The functional $Z$ satisfies the following nonlinear equation
\[
Z(r_1,r_2,y|u)=u(r_1,r_0)e^{2y\omega(r_{10})}+\]\beq
\frac{g^2N_c}{8\pi^3}\int_0^y dy'e^{2\omega(r_{10})(y-y')}\int d^2r_2
\frac{r_{10}^2}{r_{12}^2r_{20}^2}Z(r_1,r_2,y'|u)Z(r_2,r_0,y'|u),
\eeq
where $r_{10}=r_1-r_0$ etc and $\omega(r)$ is the gluon trajectory
$\omega(q)$ in which the momentum $q$ is substituted by $r$:
\beq
\omega(r)=-\frac{g^2N_c}{4\pi^3}\ln\frac{r}{\epsilon}
\eeq
with $\epsilon$ the cutoff at small $r$ (in the ultraviolet).
The functional $Z$ is normalized according to
\beq
Z(r_1,r_0,y|u)_{u=1}=1.
\eeq
The $k$-fold inclusive dipole density is given by a
$k$-fold derivative of $D$ with respect to $u$ at $u=1$:
\beq
n_k(r_1,r_0,y;\rho_1,...\rho_k)=\frac{1}{k!}\frac{\delta^kZ}
{\delta u(\rho_1)....\delta u(\rho_k)}_{u=1}.
\eeq
At this point we make our first comment as to the
comparison with [10]. Our form of the functional equation
for $Z$ is the same as in the original pater of A.H.Mueller [1]
and in [10], except for a
slightly different choice of dipole coordinates and for the order of
arguments in the 2nd $Z$ in the non-linear term: their form would
correspond to $Z(r_0,r_2,y'|u)$. Comparing with the BFKL equation
for the single dipole density, one can verify that our choice is
better. However this point is irrelevant for the following, since
in the interaction with the nucleus only even functions of $r_{10}$
appear.  

The dipoles are to interact with the nucleus
target with a zero transferred momentum. 
If the dimension of the dipole is smaller than the internucleon
distance in the nucleus then it will interact with a single nucleon
as a whole. This picture lies at the basis of the standard fan diagram
approach, where each pomeron finally interacts with a single nucleon.
We used precisely this picture in our derivation of Eq. (26) in Section 2.
In this picture the only trace of the nucleus will be an additional factor
$AT(b)$ multiplying the interaction with the nucleon $\tau(r)$  (Eq. (54)) 
at a given impact parameter $b$. 
In [10] a different idea is exploited: it is assumed that each of the 
dipoles can interact with many nucleons. The latter interaction
is assumed to have an eikonal form. So the interaction with the nucleus
they consider is a two-stage one: first the projectile generates many
colour dipoles (BFKL fan diagrams) and then each dipole multiply
interacts with the nucleus {\it a-la} Glauber.
Although technically it is not difficult to take into account the final
eikonalization of the interaction, appropriately changing the single
scattering term $\tau(r)$ in (55) and the following formulas, we do not
think it is reasonable. On the one hand, as we shall see, in the
interaction with the nucleus
the densities are considerably damped at large distances as compared to
the usual BFKL
behaviour. The confinement should further restrict their spatial
dimensions. So for a nucleus with a large internucleon distance it does
not seem reasonable to assume simultaneous interaction of a dipole with
two or more nucleons. On the other hand, should such interactions be
really important, one cannot expect to correctly describe 
the interaction with the nucleus of a dipole of a given (and fixed)
dimension by the Glauber formula. Note that the expression used in [9]
for it does not correspond to the BFKL picture for the scattering on a
single nucleon.
So we take $\tau$ as given by (55) and this is the main difference
between our derivation and that of [10].

With a chosen $\tau$, the interaction with the nucleus will be described
by
densities
\beq
\nu_k(r_1,r_0,y)=\int\prod_{j=1}^k\Big(d^4\rho_j \tau(\rho_j)AT(b)\Big)
n_k(r_1,r_0,y;\rho_1,...\rho_k).
\eeq
Here  $\tau(\rho)=\tau(\rho_f-\rho_i)$ depends only
on the
dipole length  and is given by (55). 

Differentiating (57) and using (61) one easily obtains
\[
\nu_1(r_1,r_0,y)=AT(b)\tau(r_{10})
e^{2y\omega(r_{10})}+\]\beq
\frac{g^2N_c}{8\pi^3}\int_0^y dy'e^{2\omega(r_{10})(y-y')}\int d^2r_2
\frac{r_{10}^2}{r_{12}^2r_{20}^2}
[\nu_1(r_1,r_2,y')+\nu_1(r_2,r_0,y')]
\eeq
and for $k>1$
\[
\nu_k(r_1,r_0,y)=
\frac{g^2N_c}{8\pi^3}\int_0^y dy'e^{2\omega(r_{10})(y-y')}\int d^2r_2
\frac{r_{10}^2}{r_{12}^2r_{20}^2}
[\nu_k(r_1,r_2,y')+\nu_k(r_2,r_0,y')\]\beq
+\sum_{j=1}^{k-1}
\nu_j(r_1,r_2,y')\nu_{k-j}(r_2,r_0,y')].
\eeq
We suppress the evident dependence on the impact parameter $b$.
From the structure of the
equations and the form of the inhomogeneous term it  follows
that the densities $\nu(r_1,r_0,y)$ depend only on the initial dipole length
$r_{10}$. As a result, the two terms separated from the sum over $j$ on the
right-hand side give the same contribution. One then finally finds
equations
\beq
\nu_1(r_{10},y)=AT(b)\tau(r_{10})
e^{2y\omega(r_{10})}+
\frac{g^2N_c}{4\pi^3}\int_0^y dy'e^{2\omega(r_{10})(y-y')}\int d^2r_2
\frac{r_{10}^2}{r_{12}^2r_{20}^2}
\nu_1(r_{20},y')
\eeq
and for $k>1$
\[
\nu_k(r_{10},y)=
\frac{g^2N_c}{4\pi^3}\int_0^y dy'e^{2\omega(r_{10})(y-y')}\int d^2r_2
\frac{r_{10}^2}{r_{12}^2r_{20}^2}
\nu_k(r_{20},y')+\]\beq
\frac{g^2N_c}{8\pi^3}\int_0^y dy'e^{2\omega(r_{10})(y-y')}\int d^2r_2
\frac{r_{10}^2}{r_{12}^2r_{20}^2}
\sum_{j=1}^{k-1}\nu_j(r_{12},y')\nu_{k-j}(r_{20},y').
\eeq

The total forward scattering amplitude on the nucleus will be given
by the expression (54) in which the single dipole interaction
$\int d^2r_1n_1(r,r_1,y)\tau(r_1)$ is substituted by the sum of
all multidipole interactions $\sum_k\nu_k(r)$. Presenting the
amplitude in the form (10) 
we have
\beq
\Phi(r,y)=-\sum_{k=1}\nu_k(r,y).
\eeq
Summing  Eqs. (64) and (65) over $k$ we obtain an equation
for $\Phi$:
\[
\Phi(r_{10},y)=
-AT(b)\tau(r_{10})e^{2y\omega(r_{10})}+
\frac{g^2N_c}{4\pi^3}\int_0^y dy'e^{2\omega(r_{10})(y-y')}\int d^2r_2
\frac{r_{10}^2}{r_{12}^2r_{20}^2}
\Phi(r_{20},y')-\]\beq
\frac{g^2N_c}{8\pi^3}\int_0^y dy'e^{2\omega(r_{10})(y-y')}\int d^2r_2
\frac{r_{10}^2}{r_{12}^2r_{20}^2}
\Phi(r_{12},y')\Phi(r_{20},y').
\eeq

Comparing this equation with the one derived in [10], 
apart from a different inhomogeneous term, which was discussed
earlier, we find difference in the spatial arguments of the function
$\Phi$. Our $\Phi$ depends only on one such argument: the dipole dimension
$r_{12}$. The equivalent Kovhegov's function $N$ depends on two spatial
arguments: it depends not only on the dipole dimension but also on its
center-of-mass coordinate $b_0$, not to be confused with the impact
parameter $b$ which does not enter  his equation at all. The dependence
on $b_0$ should be governed by the inhomogeneous term, which seems to 
be independent of $b_0$ (Eq. (6a) of [10]). Then the dependence of $N$ on 
the 2nd argument seems to disappear and  Kovchegov's equation 
coincides with (67). However this contradicts his initial formula 
(Eq. (4) of [10]) in which one integrates over all $b_0$. Modulo all these 
(small) inconsistencies and a different inhomogeneous term,
our Eq. (67) coincides with Kovchegov's.

\section{References}
\noi
1. A.Mueller, Nucl. Phys.,{\bf B415} (1994) 373.\\
2. A.Mueller and B.Patel, Nucl. Phys.,{\bf B425} (1994) 471.\\
3. M.A.Braun and G.P.Vacca, Eur. Phys. J {\bf C6} (1999) 147.\\
4. R.Peschanski, Phys. Lett. {\bf B409} (1997) 491.\\
5. A.Schwimmer, Nucl. Phys. B94 (1975)445.\\
6. L.V.Gribov, E.M.Levin an M.G.Ryskin, Nucl. Phys. {\bf 188} (1981) 555;
Phys. Rep. {\bf 100} (1983) 1.\\
7. J.Bartels and M.Wuesthoff, Z.Phys., {\bf C66} (1995) 157.\\
8. M.A.Braun, Eur. Phys. J {\bf C6} (1999) 321.\\
9. I.Balitsky, hep-ph/9706411; Nucl. Phys. {\bf B463} (1996) 99.\\
10. Yu. Kovchegov, Phys. Rev {\bf D60} (1999) 034008.\\
11 Yu. Kovchegov, preprint CERN-TH/99-166 (hep-ph/9905214).\\
12.E.Levin and K.Tuchin, preprint DESY 99-108, TAUP 2592-99
(hep-ph/9908317).\\
13. L.N.Lipatov in: "Perturbative QCD", Ed. A.H.Mueller, World Sci.,
Singapore (1989) 411.\\
14. M.A.Braun, Eur. Phys. J {\bf C6} (1999) 343.\\
15. N.N.Nikolaev and B.V.Zakharov, Z.Phys. {\bf C64} (1994) 631.\\
16. A.Mueller hep-ph/9904404; 9906322; 9902302.\\

\section{Figure captions}
\noi
Fig. 1. The equation summing fan diagrams. Lines represent
pomerons.\\
 Fig. 2. $\phi$ as a function of momentum at different $y$ (in units
$\pi/\alpha_sN$) for central collisions on lead ($B=0.12$, Eq.(359)).\\
 Fig. 3. $\phi$ as a function of momentum at different $y$ (in units
$\pi/\alpha_sN$) for peripheral collisions on lead ($B=0.02$, Eq.(35)).\\
 Fig. 4. Cross-section $\sigma_L$ (with $e^2\rightarrow 1$) as a
function of $y$ (in units $\pi/\alpha_sN$) for different $Q^2$.\\ 
 Fig. 5. Cross-section $\sigma_L$ (with $e^2\rightarrow 1$) as a
function of $Q^2$ for different $y$ (in units $\pi/\alpha_sN$).\\
Fig. 6. The structure function of lead  $F_{2A}(x,Q^2)$ as a
function of $x$ for different $Q^2$.\\
Fig. 7. The structure function of lead  $F_{2A}(x,Q^2)$ as a
function of $Q^2$ for different $x$.\\
Fig. 8. The gluon density (in units $\pi^2N/2\alpha_s$) as a function of
momentum at different rapidities $\tilde{y}$ for central collisions
on lead ($B=0.12$, Eq. (35)).\\                        
Fig. 9. The gluon density (in units $\pi^2N/2\alpha_s$) as a function of
momentum at different rapidities $\tilde{y}$ for peripheral collisions
on lead ($B=0.02$, Eq. (35)). 

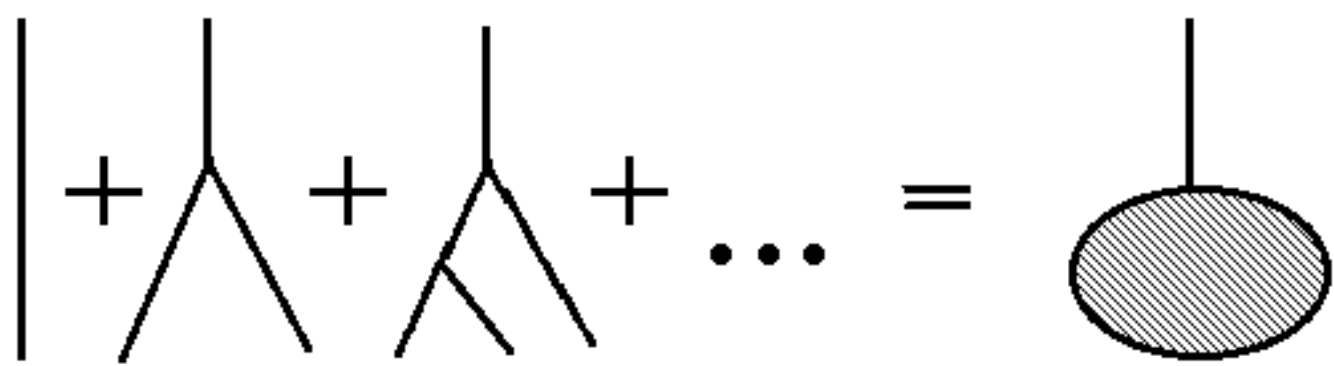
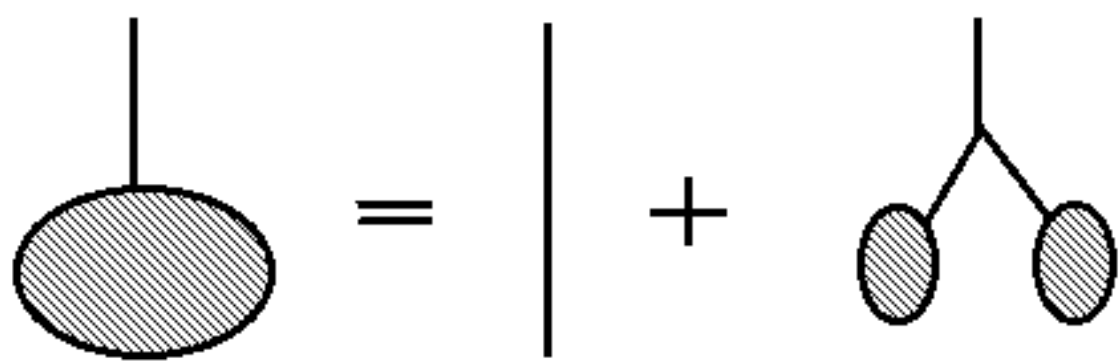
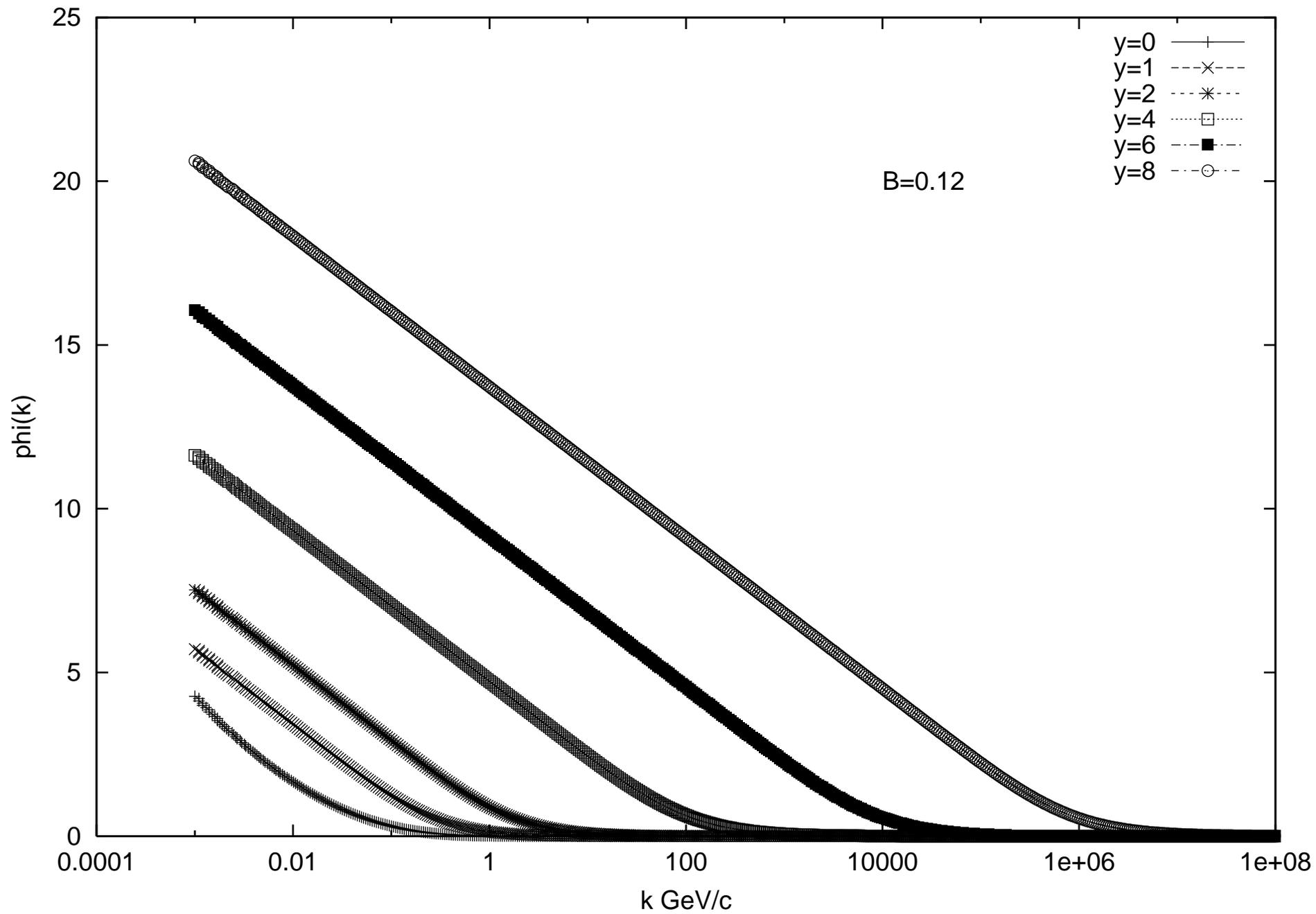
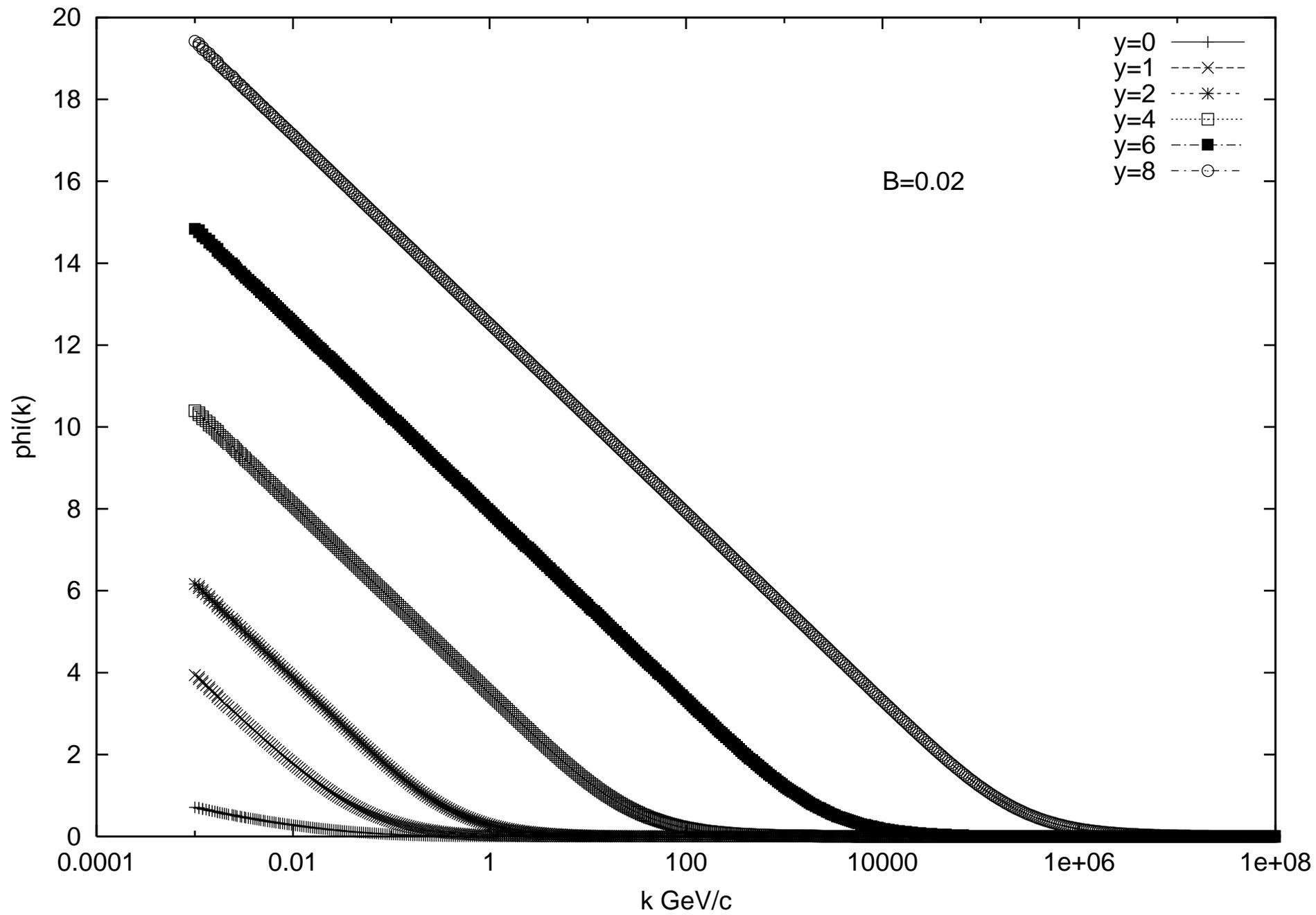
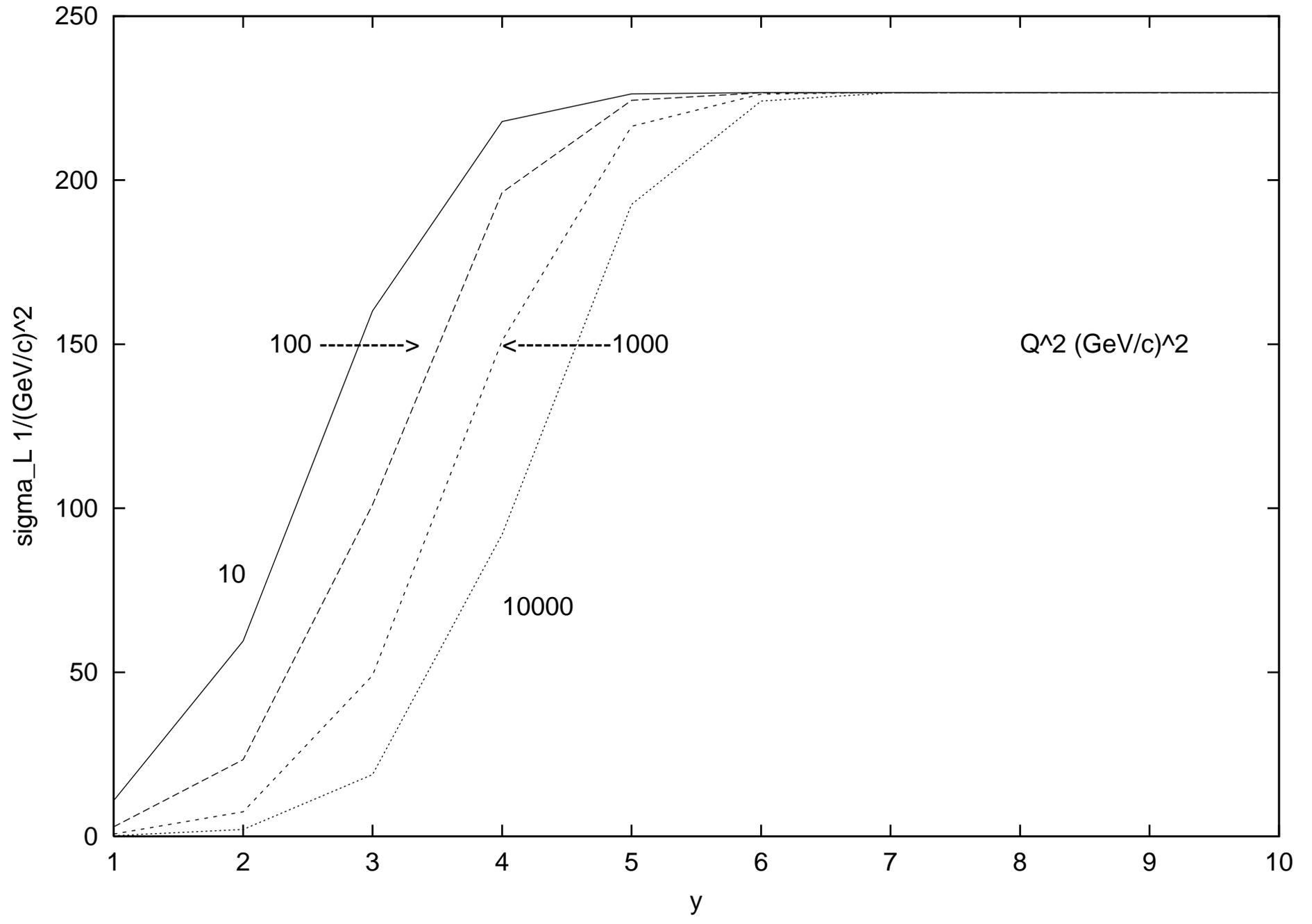
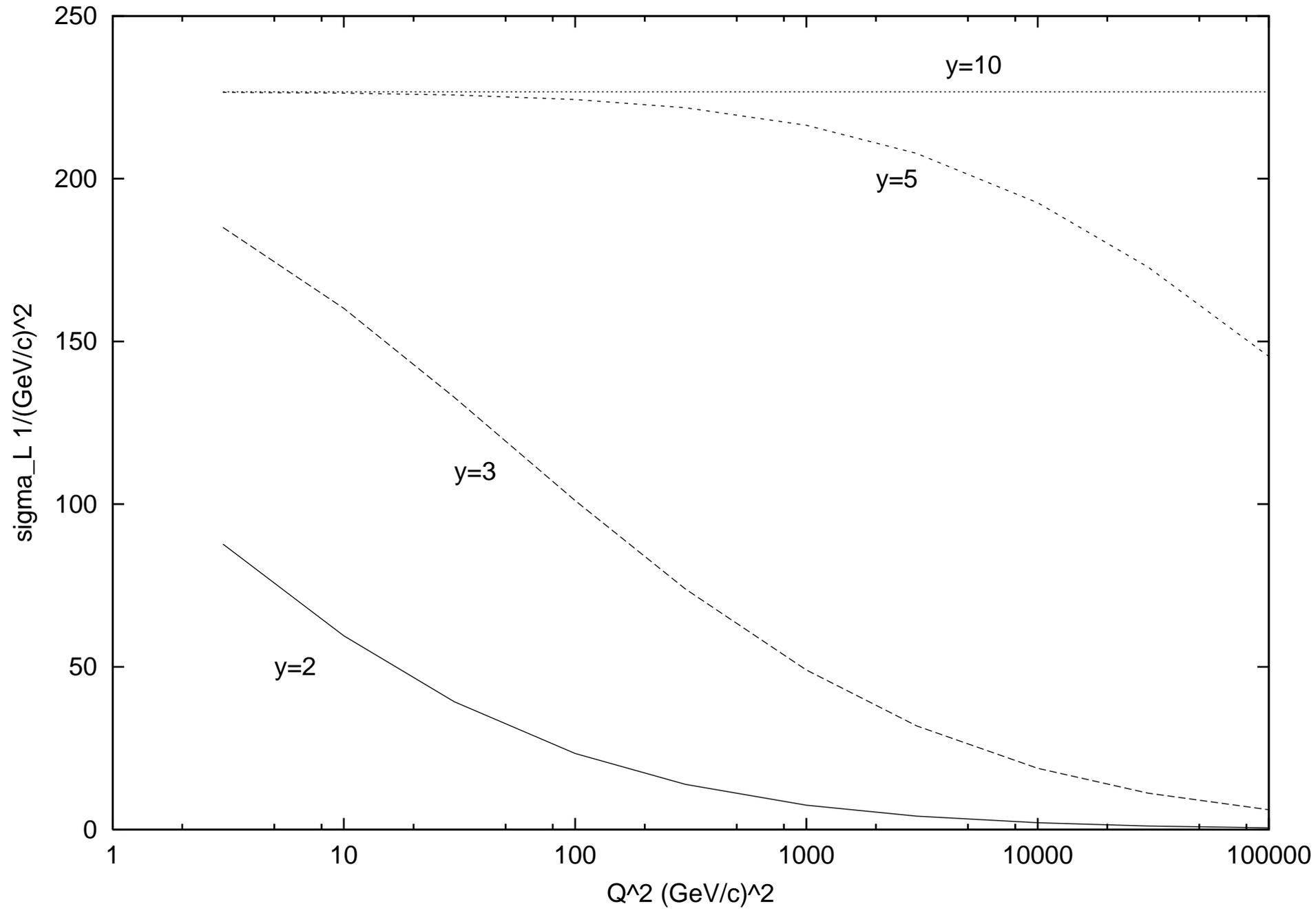
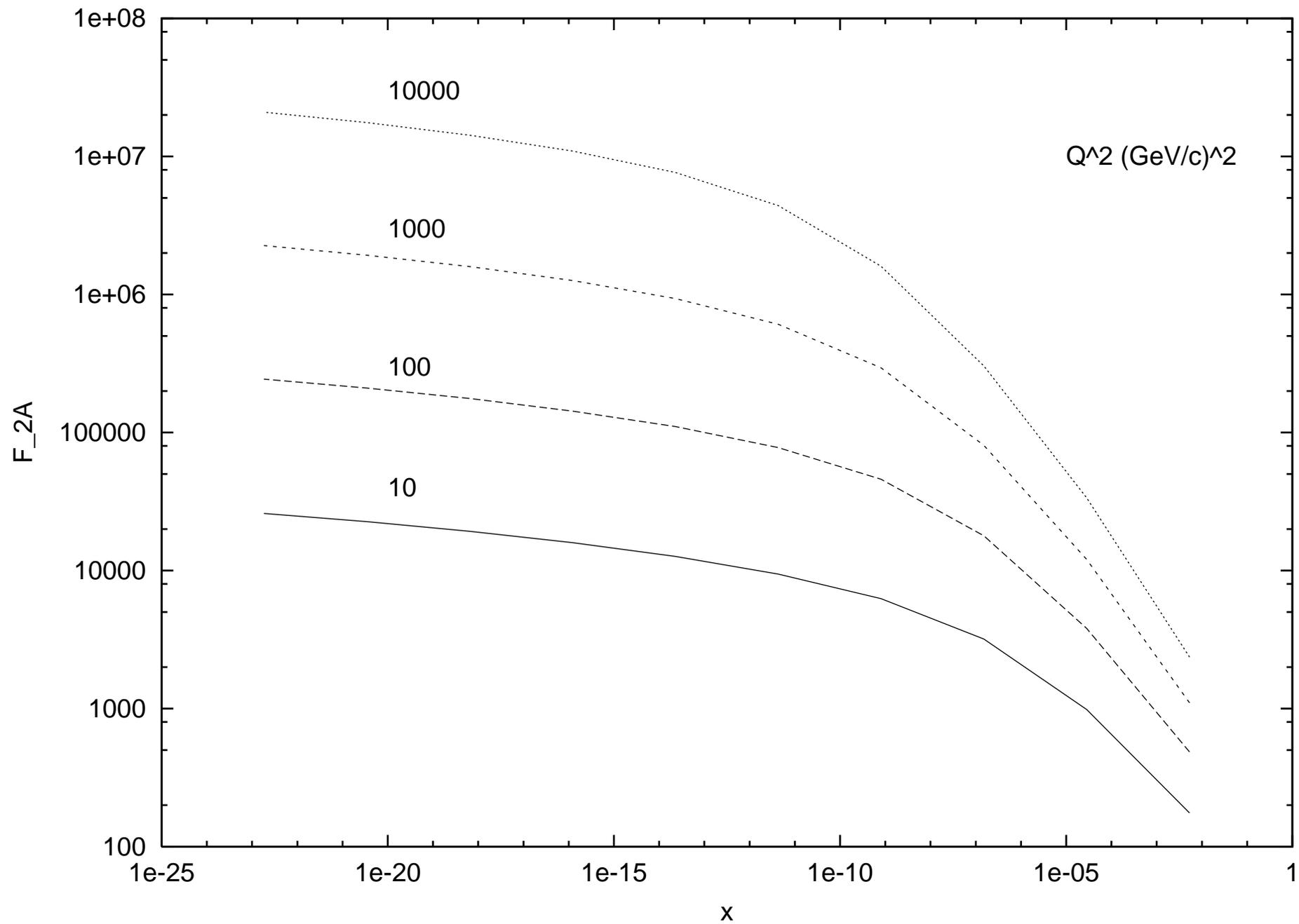
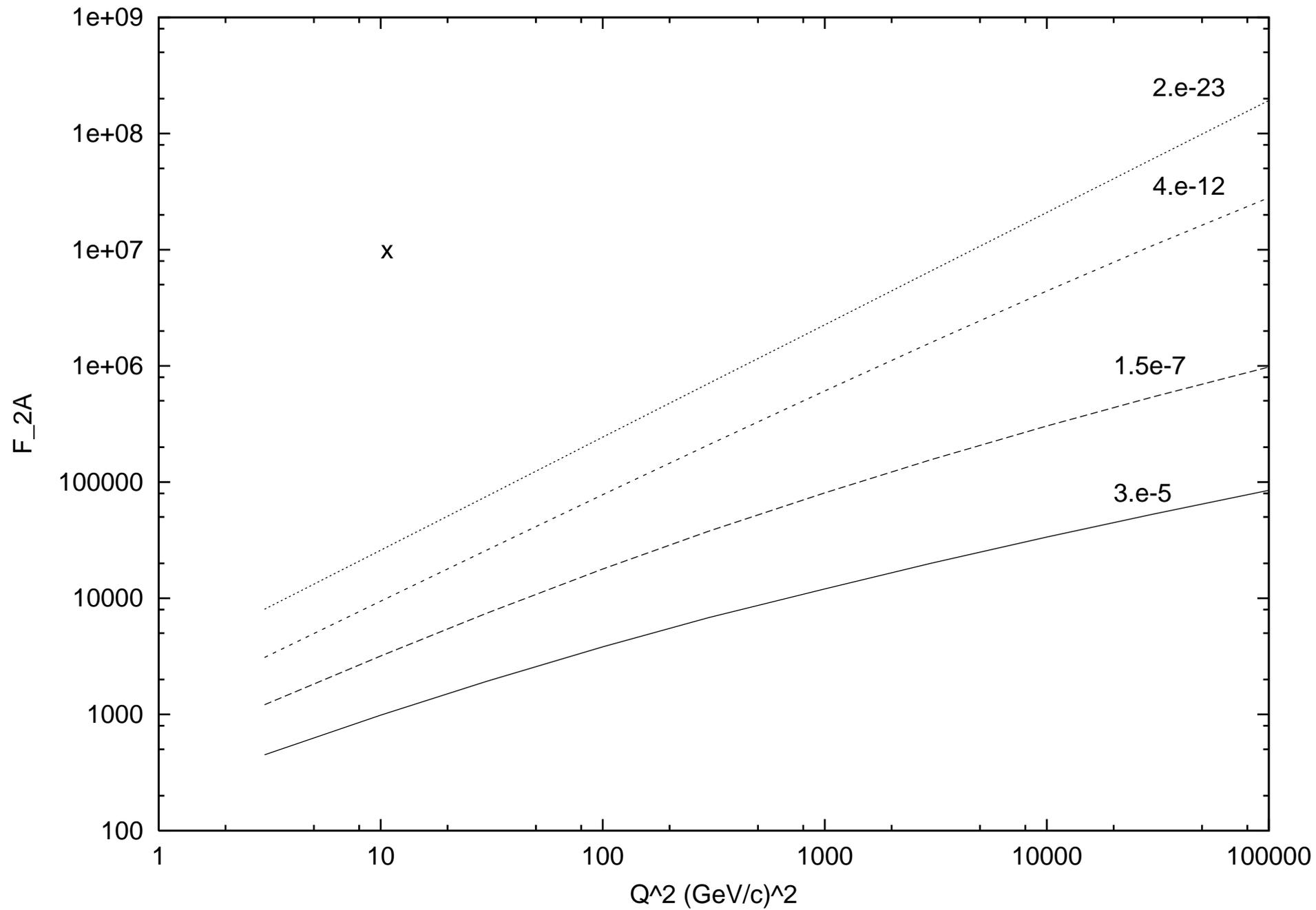
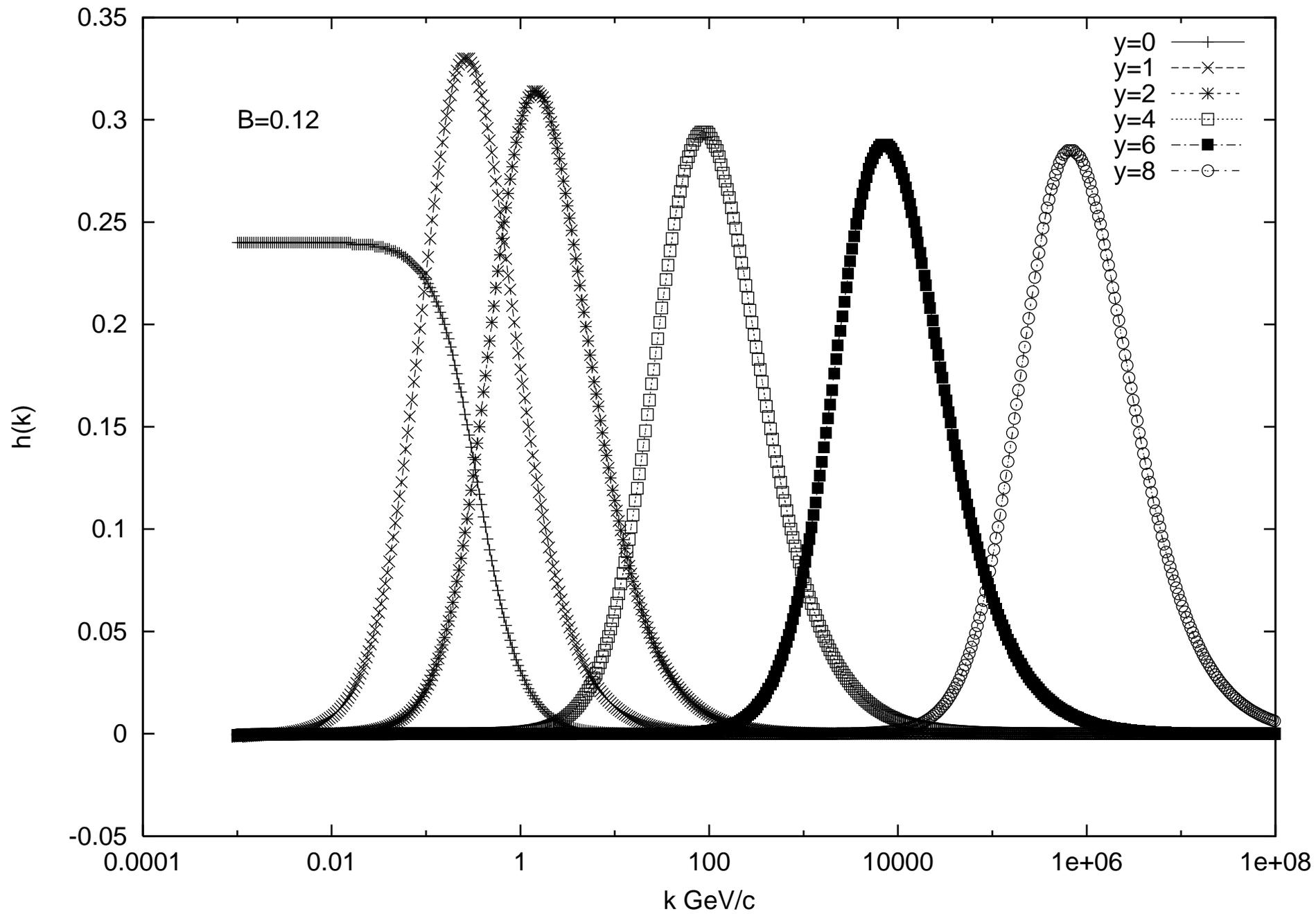
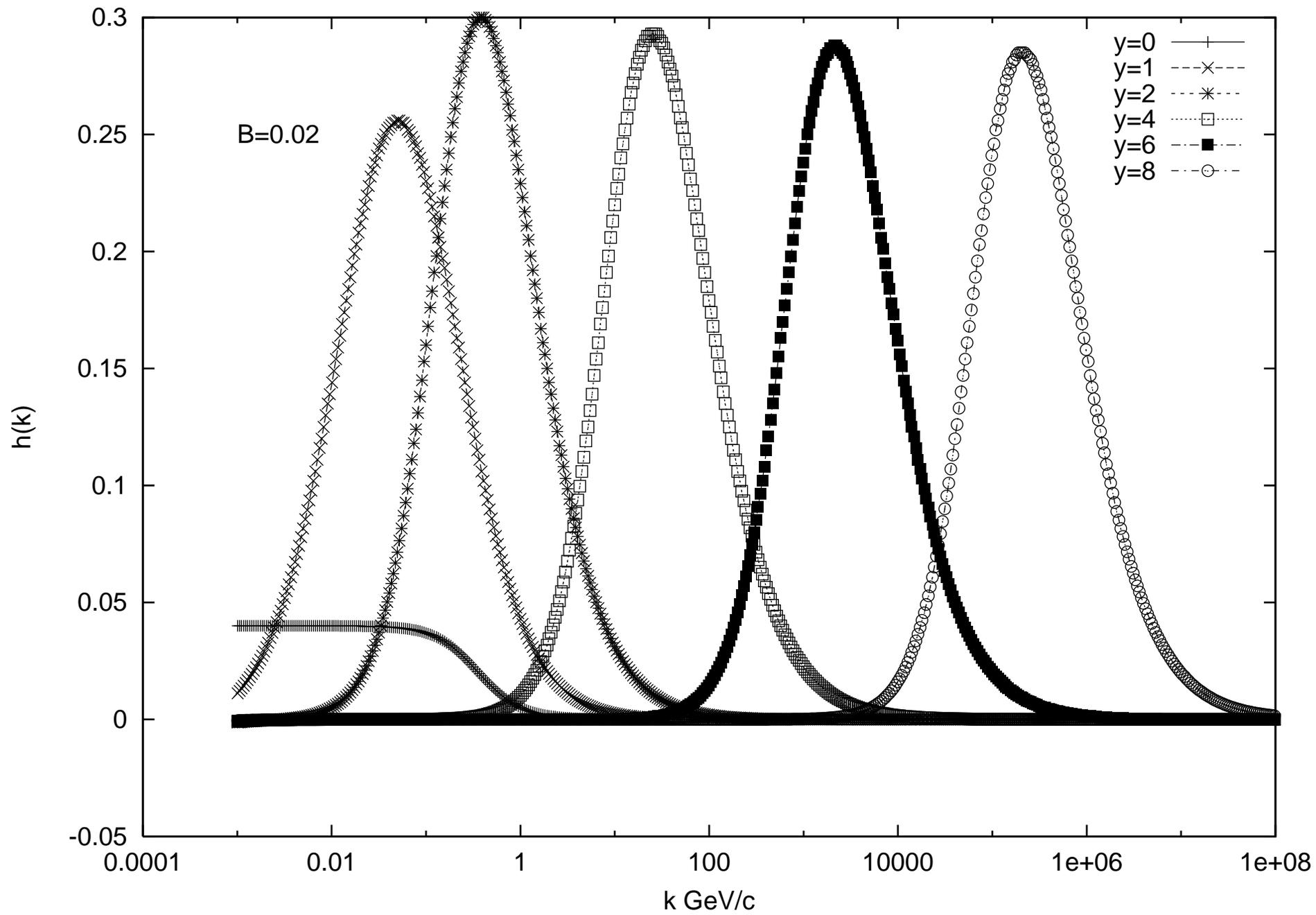
\end{document}